\documentclass[a4paper,fleqn,usenatbib]{mnras}
\usepackage{amssymb,amsmath,graphicx,psfrag,mathtools}
\usepackage[T1]{fontenc} %MNRAS
\usepackage{ae,aecompl} %MNRAS
\usepackage{url}
\usepackage{revsymb}
\hypersetup{draft} %TO AVOID ERROR: \PDFENDLINK ENDED UP IN DIFFERENT
\title[FRB 190520B]{FRB 190520B---A FRB in a Young Supernova Remnant?
}
\author[J. I. Katz]{
J. I. Katz,$^{1}$\thanks{E-mail katz@wuphys.wustl.edu} %MNRAS
\\
$^{1}$Department of Physics and McDonnell Center for the Space Sciences,
Washington University, St. Louis, Mo. 63130 USA %MNRAS
}
\date{Accepted XXX.  Received YYY; in original form ZZZ} %MNRAS
\pubyear{2021} %MNRAS
\date{\today}
\begin{document} %MNRAS
%\psfrag{theta}{$\theta$}  Works only on COMPLETE strings
\label{firstpage} %MNRAS
\pagerange{\pageref{firstpage}--\pageref{lastpage}} %MNRAS
\maketitle %MNRAS
\begin{abstract}
	FRB 190520B, a repeating FRB near-twin of FRB 121102, was discovered
	\citep{N21} to have a dispersion measure excess over the
	intergalactic and Galactic contributions of about 900 pc-cm$^{-3}$,
	attributable to its host galaxy or near-source environment.  This
	excess varies on a time scale of $\sim 30\,$y and might be explained
	by a supernova remnant no more than a few decades old.  A magnetic
	field in equipartition with the remnant's expansion would be $\sim
	1\,\text{G}$.  {I suggest the use of baseband data to measure
	very large rotation measures.}
\end{abstract}
 %MNRAS
\section{Introduction}
The recently discovered \citep{N21} FRB 190520B is, in many respects, a
near-twin of FRB 121102.  Both are identified with dwarf star-forming
galaxies with similar redshifts (0.241 and 0.193, respectively), repeat
frequently, and are accompanied by steady persistent radio sources (PRS) of
similar strength ($\sim 3 \times 10^{29}$ erg/(s-Hz) at GHz frequencies).
Neither's bursts been found to repeat periodically \citep{Z18,A21,N21},
although the activity of FRB 121102 is modulated with a 160 d period
\citep{R20,C21}.  {Insufficient data exist to decide whether FRB 190520B
is similarly modulated.}

FRB 190520B is unique among FRB in having a large contribution to its
dispersion measure (DM) in addition to the known or estimated intergalactic
and Galactic contributions, that is far in excess of their uncertainties
(Fig.~3 of \citet{N21}).  Other FRB are suspected of having such excess DM,
but they are an order of magnitude smaller and might be attributed to
underestimates of the intergalactic or Galactic contributions.  The excess
DM of FRB 190520B of $\approx 900\,$pc-cm$^{-3}$ ($\approx 2.7 \times
10^{21}$ cm$^{-2}$)\footnote{The uncertainty of this value is discussed in
detail by \citet{N21} and is unlikely to much exceed $\pm 10$\%.  The
conclusions presented here are insensitive to uncertainties of this
magnitude, so the uncertainty range is not propagated into the numerical
estimates.} is therefore attributed to matter cosmologically local to the
FRB: an interstellar cloud in the host galaxy, its halo (unlikely for the
dwarf galaxy host) or the immediate environment of the FRB.

This paper discusses these hypotheses, rejects the interstellar cloud
(Sec.~\ref{IS}; an appendix discusses the case of a cloud supported by
thermal pressure) and attributes the excess DM to a young dense supernova
remnant (SNR) enveloping the FRB (Sec.~\ref{SNR}).  If the observed mean
decrease of the DM (ED Fig.~8 of \citet{N21}) is attributed to the expansion
of a SNR \citep{K16,MM18,PG18}, its age is bounded as $\lesssim 30\,$y.

However, FRB 121102, whose DM is \emph{increasing} \citep{H20}, is not
consistent with that picture.  The rapid fluctuations of the DM of FRB
190520B, if not the result of intra-burst frequency drift, must be
attributed to turbulent motion of clumps or filaments within the SNR, as
discussed \citep{K21} for FRB 121102.  This casts doubt on the explanation
of the mean decrease, whose inference depends on a single datum, as a
consequence of SNR expansion, {and makes it difficult to relate this
to global properties of a putative SNR}.
\section{The Dispersing Cloud}
\label{IS}
\subsection{Parameters}
An assumption of fully ionized pure hydrogen is an adequate approximation.
From the measured DM the density $n$ and size $r$ of the cloud can be found
as functions of its mass $M$:
\begin{equation}
	{4 \pi \over 3} n r^3 m_p  = M,
\end{equation}
where $m_p$ is the mass of a proton.  Using the definition $\text{DM} \equiv
nr$
\begin{equation}
	\label{r}
	r = \sqrt{{3 \over 4 \pi}{M \over m_p}} \text{DM}^{-1/2}
	\approx 3 \times 10^{17} \sqrt{M \over M_\odot}\,\text{cm}
\end{equation}
and
\begin{equation}
	\label{n}
	n = \sqrt{{4 \pi \over 3}{m_p \over M}} \text{DM}^{3/2}
	\approx 8 \times 10^3 \sqrt{M_\odot \over M}\,\text{cm}^{-3}.
\end{equation}
\subsection{Emission Measure}
{Eqs.~\ref{r} and \ref{n}} may be combined to calculate the emission
measure EM if the source region is homogeneous
\begin{equation}
	\label{EM}
	\text{EM} \equiv n^2 r \approx 7 \times 10^6 \sqrt{M_\odot \over M}\
	\text{pc-cm}^{-6}.
\end{equation}

{\citet{N21} report the H$\alpha$ flux from the host galaxy of FRB
190520B.  The H$\alpha$ source is unresolved, so in order to obtain a
value for EM they must assume a solid angle for the source.  Taking this to
be the resolution element 0.25 arc-sec$^2$, they found the H$\alpha$
surface brightness $S = 503 \pm 14\,$Rayleigh (denoted $R$).}

Comparison to the value inferred from the H$\alpha$ line\footnote{The
numerical result in Eq.~6 of arXiv:2110.07418v1 does not include the factor
of $(1+z)^4$.} $\text{EM} = 3280 T_4^{0.9}\,$pc-cm$^{-6} (S/500 R)= 1.01
\times 10^{22} T_4^{0.9}\,$cm$^{-5} (S/500 R)$ \citep{N21} indicates either
a cloud of $M \sim 6 \times 10^6 (S/500 R)^{-2} T_4^{-1.8} M_\odot$ or that
the H$\alpha$ line is not produced by the same matter that provides the DM.
{The sensitive dependence of $M$ on the uncertain $S$ limits the
significance of this estimate.}

Alternatively, Eq.~\ref{EM} may be inverted to estimate $r$, using the EM
inferred from the H$\alpha$ line:
\begin{equation}
	r = {\text{DM}^2 \over \text{EM}} \approx 250\ T_4^{-0.9}
	(S/500 R)^{-1} \text{pc},
\end{equation}
implying, from Eqs.~\ref{r} and \ref{n},
\begin{equation}
	M \sim 6 \times 10^6 T_4^{-1.8} (S/500 R)^{-2}\ M_\odot
\end{equation}
and
\begin{equation}
	n \sim 3\ T_4^{0.9} (S/500 R) \text{cm}^{-3}.
\end{equation}
{If $S \sim 500 R$} these are plausible values for the mass and density
of ionized gas in a star-forming galaxy.  {However, they are very
sensitive to $S$ and hence to the unknown angular size of the H$\alpha$
source.  Large $n$ and smaller $M$ are consistent with the observations if
$S$ is large.  H$\alpha$ emission from very dense clouds is limited by
self-absorption, invalidating the assumed relation \citep{N21} between $S$
and EM.}

The value of $r$ inferred from the EM is inconsistent with the observed rapid
variation of DM unless $S \gg 500 R$, which would require that the H$\alpha$
source be very under-resolved.  The paradox may be resolved in at least two
ways:
\begin{enumerate}
	\item Most of the DM of the FRB is produced in the cloud that emits
		the H$\alpha$ radiation, but the variable part of the DM is
		produced by a much smaller cloud whose DM is undetermined
		(but must be at least as large as the amplitude of
		variation, $\Delta\text{DM} \sim 30\,\text{pc-cm}^{-3}$);
	\item The cloud that produces the H$\alpha$ radiation is not the
		source of the {host galaxy contribution to the} DM of
		the FRB, either because it is not on the line of sight, or
		because its (undetermined) DM is much less than {900
		pc-cm$^{-3}$.  This is the most economical explanation: the
		H$\alpha$ source is a large, low density, interstellar
		cloud unrelated to the FRB.}
\end{enumerate}
If the second hypothesis is correct, the observed EM does not constrain the
source of the DM.  Because 900 pc-cm$^{-3}$ exceeds the DM of known galactic
clouds, another source, {such as a SNR,} must be sought.
\section{Supernova Remnant?}
\label{SNR}
Observations (ED Fig.~8 of \citet{N21}) indicate that the DM of FRB 190520B
is declining at a mean rate of about 0.1 pc-cm$^{-3}$/d.  Comparison
to the measured excess DM of 900 pc-cm$^{-3}$ indicates a characteristic decay
time $\sim 10^4\,\text{d} \approx 30\,$y.  Both the sign and rate are
consistent with a young expanding SNR \citep{MM18,PG18}.

In addition to the mean rate of decrease of DM fitted to data over about
1 1/2 years, the DM varied irregularly by tens of pc-cm$^{-3}$ on time
scales of a few days.  This is inexplicable as the result of a cloud of
dimensions given by Eq.~\ref{r}, {and is also inconsistent with the
systematic expansion of a SNR.  It might} result from turbulent motions of
{dense} filaments within a SNR.  This explanation has been offered
\citep{K21} for the small increase in the DM of FRB 121102.

The persistent radio sources (PRS) associated with FRB 121102 and FRB
190520B are plausibly produced by a young supernova remnant.  However,
\citet{N21} found (ED Table 1) that its flux at 3 GHz \emph{increased} by
$43 \pm 13\,\mu$Jy ($24 \pm 7\,$\%, {nominal $1\sigma$ uncertainty})
over 68 days.  Like the fluctuating DM, this is not consistent with a
simple model of an expanding SNR.  Alternative models have included an
intermediate mass black hole source of FRB \citep{K17,K19,K20}.

Combining a nominal transverse (to the line of sight to the FRB) velocity
$v = v_8 \times 10^8\,$cm/s with an observed (ED Fig.~8 of \citet{N21}) DM
fluctuation time scale $t = t_5 \times 10^5\,$s implies a transverse
spatial scale {of the structure responsible for the fluctuations of the
DM} $s = vt = 10^{13}v_8 t_5\,$cm.  {This parametrization may be
informative even if the cloud is not a SNR.}  The density
\begin{equation}
	n \sim {\Delta\text{DM} \over s} \sim {10^7 \over v_8 t_5}\
	\text{cm}^{-3},
\end{equation}
where the rapid variations $\Delta\text{DM} \sim 30\,\text{pc-cm}^{-3} \sim
10^{20}\,\text{cm}^{-2}$.
\section{Magnetic Stress and Rotation Measure}
If there is a magnetic stress comparable to the characteristic hydrodynamic
stress 
\begin{equation}
	{B^2 \over 8\pi} \sim n m_p v^2 \sim {\Delta\text{DM} m_p v \over t}
	\sim 0.1 {v_8 \over t_5}\ {\text{erg} \over \text{cm}^3}
\end{equation}
and
\begin{equation}
	B \sim 1.5 \sqrt{v_8 \over t_5}\ \text{G}.
\end{equation}
{For plausible $v_8 \sim 1$ and the observed $t_5 \sim 1$ the field
$B \sim 1\,$G.}  The rotation measure
\begin{equation}
	\text{RM} \sim {e^3 \over 2 \pi m_e^2 c^4} \Delta\text{DM}\,B \sim
	3 \times 10^7 \sqrt{v_8 \over t_5}\ {\text{radian} \over \text{m}^2}.
\end{equation}

This RM is so large that it is difficult to measure directly because of
intra-channel Faraday rotation, but could be manifested as the observed
absence of linear polarization \citep{N21}.  {It might be measurable
with high rate sampling of the baseband signal
\citep{K15,G18,L19,L20,Mi21,Mc21}.  The dispersive time delay
\begin{equation}
	t = \text{DM}{e^2 \over 2 \pi m_3 c}\nu^{-2},
\end{equation}
where DM is in units of electrons/cm$^2$.  The birefringent rotation angle
\begin{equation}
	\phi = \text{RM} c^2 \nu^{-2},
\end{equation}
where RM is in units of radian/cm$^2$.  Eliminating the frequency $\nu$,
\begin{equation}
	\phi = {\text{RM} \over \text{DM}} {2 \pi m_e c^3 \over e^2} t
\end{equation}
and the electric field in the direction of a linearly polarized feed would
have the time dependence
\begin{equation}
	E_\parallel(t) \propto F(t)\sin{(\phi + \phi_0)},
\end{equation}
where $F(t)$ is a comparatively slowly varying function describing the 
envelope of the burst intensity and $\phi_0$ a fixed phase offset.

The Fourier transform of $E_\parallel(t)$ would have an amplitude peak at the
angular frequency
\begin{equation}
	\begin{split}
		\omega_{pol} &= {\text{RM} \over \text{DM}}
		{2 \pi m_e c^3 \over e^2}\\
		&= {\text{RM} \over 10^6/\text{m}^2}
		{1000\,\text{pc-cm}^{-3} \over \text{DM}}\
		2.16 \times 10^4\ \text{s}^{-1}
	\end{split}
\end{equation}
and the transform of $|E_\parallel(t)|$ (proportional to the intensity)
would peak at $2\omega_{pol}$.  The factor $\omega_0 \equiv 2 \pi m_e
c^3/e^2 = 6.68 \times 10^{23}\,$s$^{-1}$ (the energy $\hbar \omega_0 =
2 \pi m_e c^2/\alpha$, where $\alpha$ is the fine-structure constant).
These peaks are broadened by the spectral width of $F(t)$.}
\section{Discussion}
The data may be explained if there are two clouds: a large cloud that is the
source of the (apparently steady) H$\alpha$ emission, and a small cloud that
produces the large and variable local DM.  This second cloud is naturally
interpreted as a young SNR in which the FRB is embedded.  These clouds may
be unrelated.

Unfortunately, if this explanation is correct the measured DM cannot be used
to constrain the parameters of the larger cloud, the source of the H$\alpha$
radiation.  Nor can the larger cloud's EM be used to constrain the
parameters of the source of the excess DM.

The inference of an age $\lesssim 30\,$y depends on the assumption that the
DM trend fitted in ED Fig.~8 of \citet{N21}, largely (but not entirely)
dependent on a single early datum, is really a long-term trend, and not the
result of accidental fluctuations of the rapidly varying DM.  This can soon
be tested by new data, but cannot yet be proven or disproven.

{The dispersing cloud radius Eq.~\ref{r} corresponds to an age $\approx
10 \sqrt{M/M_\odot}/v_9\,$y, consistent with the mean decrease of the DM.
However, the rapid irregular fluctuations of DM are inconsistent with a
simple expanding SNR model, and require appeal to a filamentary structure.
The $3\sigma$-significant increase of the 3 GHz flux of the persistent
source, in two months, if real, is also inconsistent and inexplicable as the
result of dense filaments of thermal gas.  Such rapid variations have not
been reported in well-studied young (but not as young as estimated here) SNR
\citep{T17}.}
\section*{Data Availability}
This theoretical study did not generate any new data.
\section*{Acknowledgements}
I thank C. Law and W. Lu for comments and discussions that have improved
this paper.

\appendix
\section{Thermally Supported Clouds}
If a homogeneous spherical cloud is supported by thermal pressure at a
temperature $T$ then there is an approximate relation among its mass $M$,
radius $r$, atomic density $n$ and temperature
\begin{equation}
	\label{TS}
	{4 \pi \over 3}{Gnr^3 m_p^2 \over 2r} \approx {GMm_p \over 2r}
	\approx k_B T,
\end{equation}
where we approximate the mean mass per particle as $m_p/2$.  The
conclusion is sufficiently robust that we need no better approximation.

Using the definition of DM, Eq.~\ref{TS} implies
\begin{equation}
	\label{TS2}
	r = {3 \over 2\pi}{k_B T \over G m_p^2 \text{DM}} \approx 425\ T_4\
	\text{pc},
\end{equation}
where $T_4 \equiv T/(10^4\,\text{K})$.  This is inconsistent with the rapid
variations of DM shown in ED Fig.~8 of \citet{N21}, so the hypothesis that
the local contribution to the DM of FRB 190520B is produced by a
pressure-supported interstellar cloud can be definitively rejected. 

If the cloud were supported by turbulent motion rather than by thermal
pressure, its lifetime would be short, either because these motions would
disrupt it or because they would thermalize.  The gas in our Galaxy is
mostly supported by organized rotation, rather than either thermal pressure
or turbulence, yet no line of sight other than to the Galactic center has a
DM within an order of magnitude of that of the near-source contribution to
the DM of FRB 190520B.
\label{lastpage} %MNRAS
\end{document}